\documentclass[twocolumn]{aastex631}
\usepackage{textcomp}

\usepackage{gensymb}
\usepackage{amsmath}

\newcommand{\om}{$\omega$~Cen}

\usepackage{multirow} 

\usepackage{hyperref}

\begin{document}

\title{No evidence for accretion around the intermediate-mass black hole in Omega Centauri}

\author[0000-0003-0851-7082]{Angiraben D. Mahida}
\affiliation{International Centre for Radio Astronomy Research -- Curtin University, GPO Box U1987, Perth, WA 6845, Australia}

\author[0000-0001-9261-1738]{Arash Bahramian}
\affiliation{International Centre for Radio Astronomy Research -- Curtin University, GPO Box U1987, Perth, WA 6845, Australia}

\author[0000-0003-3124-2814]{James C.A. Miller Jones}
\affiliation{International Centre for Radio Astronomy Research -- Curtin University, GPO Box U1987, Perth, WA 6845, Australia}

\author[0000-0001-9261-1738]{Susmita Sett}
\affiliation{International Centre for Radio Astronomy Research -- Curtin University, GPO Box U1987, Perth, WA 6845, Australia}

\author[0000-0002-8532-4025]{Kristen Dage}
\affiliation{International Centre for Radio Astronomy Research -- Curtin University, GPO Box U1987, Perth, WA 6845, Australia}

\author[0000-0002-1468-9668]{Jay Strader}
\affiliation{Center for Data Intensive and Time Domain Astronomy, Department of Physics and Astronomy, Michigan State University, East Lansing, MI 48824, USA}

\author[0000-0002-2801-766X]{Timothy J. Galvin}
\affiliation{CSIRO Space \& Astronomy, PO Box 1130, Bentley WA 6102, Australia}

\author[0000-0003-0725-8330]{Alessandro Paduano}
\affiliation{International Centre for Radio Astronomy Research -- Curtin University, GPO Box U1987, Perth, WA 6845, Australia}





\begin{abstract}

For over a decade, both theoretical predictions and observational studies have suggested that $\omega$~Centauri (\om), the most massive Milky Way globular cluster, might harbor an intermediate-mass black hole (IMBH). Recently, identification of fast-moving stars in the core of \om\ provided the strongest evidence to date for the presence of such an IMBH. One of the key questions in the study of IMBHs is their accretion efficiency, which determines their radio and X-ray signatures. We investigate the accretion signature of the IMBH in \om\ with ultra-deep radio continuum observations of the central region of the cluster. Using approximately 170\,hours of Australia Telescope Compact Array observations, we achieve a root mean square noise of 1.1~$\mu$Jy at 7.25\,GHz, making this the most sensitive radio image of the cluster to date. We detect no radio emission at any of the proposed centers of the cluster, imposing stringent constraints on the presence of an accreting IMBH in \om. Considering the fundamental plane of black hole activity, our findings indicate that the accretion efficiency around the black hole is exceptionally low (with a conservative 3-$\sigma$ upper limit of $\epsilon \lesssim 4\times10^{-3}$). 

\end{abstract}



\section{Introduction} \label{sec:intro}

Over five decades of X-ray observations \citep{1962PhRvL...9..439G,2006ARA&A..44...49R}, together with the past decade of gravitational wave detections \citep{2016PhRvL.116f1102A,2025arXiv250818082T}, have led to the identification of a population of stellar-mass black holes \citep[sMBHs, \(\lesssim 200~M_{\odot}\); e.g.,][]{1999ApJ...522..413F, 2001Natur.413..139M} to complement the well-established population of supermassive black holes \citep[SMBHs, \(\gtrsim 10^5~M_{\odot}\); e.g.,][]{1995ARA&A..33..581K, 2013ARA&A..51..511K}. By contrast, we have comparatively little observational evidence \citep[e.g.,][]{2015ApJ...809L..14B,2019ApJ...872..104N} for a population of intermediate-mass black holes (IMBHs, with masses between \(\sim200\) and \(10^5\) M$_\odot$), which are the missing link in our understanding of black hole evolution. IMBHs observed today offer critical insights into the origin and early growth channels of SMBHs \citep{2005MNRAS.358..913V,2013ARA&A..51..511K,2018Natur.553..473B,2020ARA&A..58..257G}.

IMBHs may form through repeated mergers in dense star clusters, as shown in the simulations by \cite{2002ApJ...576..899P,2024MNRAS.531.3770R,2025MNRAS.543.2130R} or via direct gas collapse \citep{2006MNRAS.371.1813L,2008MNRAS.387.1649B,2014Sci...345.1330A,2019PhRvD.100d3027R,2021MNRAS.501.1413N}. Another possible formation channel is the collapse of population III stars into black holes ($ \geq 100$M$_\odot$), followed by mergers leading to an IMBH \citep{1998ApJ...503..505H,2002MNRAS.330..232C,2004ARA&A..42...79B,2005ApJ...634..910W}. However, even after decades of searching, the formation and presence of IMBHs still remain elusive due to the paucity of observational evidence \citep[e.g.,][]{2020ARA&A..58..257G}.

Dynamical studies and accretion models suggest that globular clusters (GCs) and dwarf galaxies are promising sites for IMBH searches, particularly given the detection of candidate IMBHs (or perhaps light SMBHs) as some dwarf galaxies' central black holes \citep[$10^6$M$_\odot$ $\lesssim$ $M$ $\lesssim 10^{10}$\,M$_\odot$,][]{2014Natur.513..398S,2015ApJ...799...98M,2019ApJ...872..104N}. Another effective technique for sensing the presence of IMBHs is through a dynamic analysis of the stars residing in the centers of GCs. If a central IMBH is present, it will leave a distinct signature on the line-of-sight velocities and proper motions of the surrounding central stars \citep{2002AJ....124.3270G, 2003AJ....125..376G}. However, these signatures could also indicate a population of mass-segregated stellar-mass compact objects \citep{2024ApJ...975..268S,2025A&A...693A.104B}. Additionally, detecting IMBHs using dynamical analyses is challenging as the sphere of influence of an IMBH could be small and, therefore, would have only a limited number of stars \citep[e.g.,][]{2010ApJ...710.1063V}.

Another method to search for evidence of IMBHs is through their accretion signatures, which generate X-ray and radio emission. X-ray emission arises from the accretion flow, while synchrotron radiation often arises from jets, producing radio continuum emission. In GCs, the winds of evolved stars continuously supply gas to the intracluster medium, and even minimal accretion by an IMBH could generate significant X-ray luminosity \citep{2006ApJ...644L..45P}. Our best tool for interpreting X-ray and radio emission from a black hole is the fundamental plane of black hole activity. This phenomenological correlation links X-ray and radio luminosities with the accreting black hole's mass \citep{2003MNRAS.345.1057M, 2004A&A...414..895F}. If the assumptions underlying the fundamental plane hold for a given system, the observed X-ray and radio luminosities can be used to estimate the black hole mass \citep[e.g.][]{2005MNRAS.356L..17M,2011ApJ...729L..25L,2018ApJ...862...16T, 2024ApJ...961...54P}. However, this correlation exhibits an intrinsic scatter with a standard deviation in the mass dimension of approximately 1 dex \citep{2019ApJ...871...80G}.

Over the past decade, numerous studies have investigated the presence of an IMBH at the center of \om\ (NGC 5139), with proposed masses spanning a wide range—from as low as $1200$\,M$\odot$ \citep[e.g.,][]{2010ApJ...710.1032A} to recent strong evidence suggesting a mass between $39,000$ and $47,000$\,M$\odot$ \citep{2024Natur.631..285H}. \om~is in some ways a unique cluster as its observational properties suggest a complex evolutionary history, straddling the line between the tidally stripped nuclear star cluster of a dwarf galaxy and a globular cluster \citep[e.g.,][]{2000A&A...362..895H,2003MNRAS.346L..11B}. It is the brightest and most massive {\citep[$3.5 \times 10^{6}M_{\odot}$;][]{2018MNRAS.478.1520B} star cluster in the Milky Way with a broad metallicity distribution and multiple stellar populations \citep{1975ApJ...201L..71F,2002ASPC..265..143F,2017MNRAS.469..800M}. Because of these properties \om~ has been often suggested to be the remnant nucleus of an accreted dwarf galaxy \citep[e.g.,][]{2003MNRAS.346L..11B,2003ApJ...589L..29T}. Numerical simulations showed that the merger of the \om\ progenitor with the Milky Way would result in stripping away the outer layers and leaving the orbitally bound nucleus, which is consistent with observations \citep{2003MNRAS.346L..11B}. Furthermore, the relatively low distance of \om\ \citep[$5494\pm61$\,pc;][]{2025ApJ...983...95H} makes it an effective target for IMBH searches. There have been several dynamical studies hinting at evidence of a putative IMBH in \om~ \citep{2008ApJ...676.1008N, 2008IAUS..246..341N, 2010ApJ...710.1063V, 2010ApJ...710.1032A, 2010ApJ...719L..60N, 2010AAS...21534006N, 2017MNRAS.464.2174B, 2024MNRAS.528.4941P, 2024Natur.631..285H}.
While \om\ has been the target of a deep X-ray survey using the Chandra X-ray observatory, no central source consistent with an IMBH has yet been identified \citep{2018MNRAS.479.2834H, 2022MNRAS.516.1788S}. Similar IMBH searches in the radio \citep{2011ApJ...729L..25L, 2018ApJ...862...16T} and infrared \citep{2025arXiv251120945C} bands have found no evidence of jets from an accreting central black hole, and N-body simulations \citep{2019MNRAS.488.5340B,2025A&A...693A.104B} have debated the claims regarding the presence of an IMBH.

In this work, we conduct the deepest ever radio study of \om\ and its center, to search for the radio signature of an accreting IMBH. In Section \ref{sec:data}, we provide the details of the data and describe the processing, in Section \ref{sec:results} we present our main results, and in Section \ref{sec:diss} we discuss the interpretation of our findings in the context of the presence of an IMBH in \om.

\section{Observation and data reduction} \label{sec:data}

\subsection{ATCA 2024 data}

We observed \om~with the Australia Telescope Compact Array (ATCA, \citealt{2011MNRAS.416..832W}) in 2024 under project code CX556. Throughout these observations, the array was in 6A and 6C configurations. The extended 6\,km array configuration was selected to maximise the spatial resolutions of radio images. Table \ref{tab:observations} provides detailed information on our observations of \om. 

Our observations were carried out in two wide intermediate frequency bands (IFs): one centered at 5.5\,GHz and the other centered at 9.0\,GHz (each IF with a bandwidth of 2048\, 1\,MHz channels). PKS 1934$-$638 was used as the primary calibrator for bandpass and flux calibration. PKS 1315$-$46 was used as the secondary calibrator for phase and amplitude calibration. Additionally, we observed PKS~0823$-$500 as a backup primary calibrator. During the observations, PKS~0823$-$500 was initially observed for approximately 10--15\,minutes after the telescope setup. Following this, we alternated between observing the source and the secondary calibrator, allowing 15\,minutes for the source and 2\,minutes for the calibrator. During poor weather conditions, the observation time for the source was reduced to 10\,minutes in each cycle. Most of the observations ended with observing the primary calibrator PKS~1934$-$638 for 10--15\,minutes.

\subsection{ATCA archival data}
\om~was targeted at least twice previously with sensitive, high-frequency observations, under projects C2158 \citep{2011ApJ...729L..25L} and C2877 \citep{2022MNRAS.513.3818T}. All other archival observations were either in lower-frequency bands focusing on pulsar searches or were from prior to the bandwidth upgrade \citep{2011MNRAS.416..832W}, when the telescope was far less sensitive \citep{2005MNRAS.356L..17M}.} The first project was carried out before the installation of new 4\,cm receivers on ATCA making the former observations (C2158; 18.3\,hours on source) less sensitive than the remaining data. The latter project (C2877) includes post-upgrade data amounting to 11.8\,hours on source. Both studies carried out observations in the 6 km configurations, and observed the source PKS 1934$-$638 as the primary calibrator and PKS 1320$-$446 as the secondary calibrator. The root mean square (rms) noise reported in previous studies was \(6.5\)~$\mu$Jy beam$^{-1}$ (6.8\,GHz) in \citet{2011ApJ...729L..25L} and \(2.9\)~$\mu$Jy beam$^{-1}$ (7.25\,GHz) in \citet{2022MNRAS.513.3818T} for the stacked 5.5- and 9.0-GHz data. Both studies concluded that no radio emission was detected at or near the center \citep{2010ApJ...710.1063V} of \om~above their observed sensitivity limit.

\begin{table}[ht!]
\centering
\caption{The ATCA observations of Omega Cen.}
\label{tab:observations}
\begin{tabular}{lcccc}
\hline
\hline
\textbf{Project} &\textbf{Date} & \textbf{Start time} & \textbf{Time on Source}  \\
                 &              & (UTC)               & (hr)  \\
\hline
\hline
CX556 & 2024-12-27 & 13:29 & 7.89  \\
& 2024-12-26 & 13:35 & 8.90  \\
& 2024-12-23 & 14:17 & 9.01  \\

& 2024-09-29 & 03:00 & 5.02  \\
& 2024-09-28 & 02:00 & 5.99  \\
& 2024-09-25 & 20:49 & 5.05  \\
& 2024-09-15 & 21:26 & 9.99  \\
& 2024-09-04 & 04:00 & 4.65  \\
& 2024-09-01 & 02:04 & 5.99  \\
& 2024-08-31 & 02:03 & 5.08  \\
& 2024-08-30 & 22:15 & 4.25  \\
& 2024-08-29 & 05:02 & 5.10  \\

& 2024-03-24 & 09:01 & 9.82  \\
& 2024-03-06 & 09:50 & 9.99  \\
& 2024-03-05 & 10:19 & 9.89  \\
& 2024-03-02 & 10:21 & 9.89  \\
& 2024-02-29 & 19:54 & 1.48  \\
& 2024-02-28 & 10:01 & 9.89  \\
& 2024-02-23 & 10:57 & 9.82  \\
& 2024-02-14 & 12:29 & 9.39  \\
\hline
C2877 & 2015-01-09 & 14:04 & 7.04  \\
& 2014-12-06 & 19:16 & 1.91  \\
& 2014-12-04 & 15:22 & 2.84  \\
\hline
C2158 & 2010-01-23 & 01:14 & 7.61  \\
&2010-01-22 & 12:31 & 10.70  \\
\hline
\hline
\end{tabular}
\tablecomments{C2158 observations were carried out prior to the installation of new receivers on ATCA.}
\end{table}

\subsection{Data processing}
We processed all data using the software \textsc{Miriad} \citep{1995ASPC...77..433S}, following the standard procedures for flagging and calibration\footnote{\url{https://www.atnf.csiro.au/computing/software/miriad/userguide/}}. After conducting bandpass, flux density and time-varying gain calibrations, the visibilities were imported into the Common Astronomy Software Application (\textsc{CASA 5.6}; \citealt{2022PASP..134k4501C})\footnote{\url{https://casaguides.nrao.edu/index.php/CASA_Guides:ATCA_Advanced_Continuum_Polarization_Tutorial_NGC612-CASA4.7}} for imaging. We used the {\tt\string tclean} task to generate images and apply Briggs weighting with robustness of 1.0 to balance image sensitivity and resolution. While imaging, we used Multi-Term Multi-Frequency Synthesis {\tt\string (mtmfs)} and {\tt\string mvc} spectral modes along with two Taylor terms {\tt\string (nterms=2)}. We used a cell size of \(0.15\)$^{\prime\prime}$ and an image size of 5760 pixels for imaging the 5.5 and 9.0\ GHz bands. The synthesized beams in the 5.5 and 9.0\ GHz stacked images of 172\,hours are $2.9^{\prime\prime}\times 1.7^{\prime\prime}$, and $1.8^{\prime\prime}\times 1.1^{\prime\prime}$, respectively. The stacked image of these two frequencies has an apparent central frequency of 7.25\,GHz, which we imaged with a cell size of \(0.15\)$^{\prime\prime}$ and an image size of 6912 pixels. The synthesized beam in the stacked image of 7.25\ GHz is $2.1^{\prime\prime}\times 1.3^{\prime\prime}$.

\section{Results} \label{sec:results}

By combining all the available ATCA data, we produced deep radio images of the cluster at both 5.5 and 9.0\,GHz, reaching noise rms values of \(1.7\)~$\mu$Jy beam$^{-1}$ and \(1.4\)~$\mu$Jy beam$^{-1}$, respectively. We also combined the data across the frequencies, making the deepest radio image of \om\, with a noise rms of \(1.1\)~$\mu$Jy beam$^{-1}$ at an apparent frequency of 7.25\,GHz (Fig.~\ref{fig:omegacen}).

The rms values reported here are measured near the phase center. In ideal experiments, it is expected that rms would decrease as the inverse square-root of accumulated exposure time. However, performing step-stacking, we found that in these observations, the rms noise scaled with exposure time with power law indices of $-0.38$ (5.5 GHz) and $-0.46$ (9.0 GHz). 

\subsection{Radio emission constraints in the Core}

Given the complexity of \om, there have been several estimates for the exact position of the cluster center. In this work, we explore the presence of radio emission close to the three recently-proposed center locations, derived by \citet{2024MNRAS.528.4941P} and \citet{2024Natur.631..285H}, through analyses that build upon and extend earlier studies by \citet{2008ApJ...676.1008N, 2010ApJ...710.1032A, 2010AAS...21534006N}. \citet{2024MNRAS.528.4941P} analysed \om~using high-resolution MUSE spectroscopy, extracting velocities for 28,108 stars. They discovered a counter-rotating core within the inner $\sim$20 arcsec. They derived the kinematic center from the counter-rotating core's rotation axis via kinemetry analysis \citep{2006MNRAS.366..787K}, and the dispersion center from the peak in the central velocity dispersion profile. The centers are offset by $\sim$10 arcsec, suggesting complex core dynamics and possibly indicating a wandering or off-center IMBH. Using HST data, \citet{2024Natur.631..285H} identified an IMBH in \om~through the detection of seven fast-moving stars with velocities exceeding the cluster's escape velocity ($v_{\rm esc}=62$ km s$^{-1}$). The IMBH location that they derived was based on proper motion analysis from 20 years of HST observations covering 1.4 million stars \citep{2024ApJ...970..192H}, which is roughly between the two proposed centers by \citep{2024MNRAS.528.4941P} and could be the center of the cluster. These center positions are listed in Table~\ref{tab:centers}.

\begin{figure*}[ht!]
\centering
\includegraphics[width=1\textwidth]{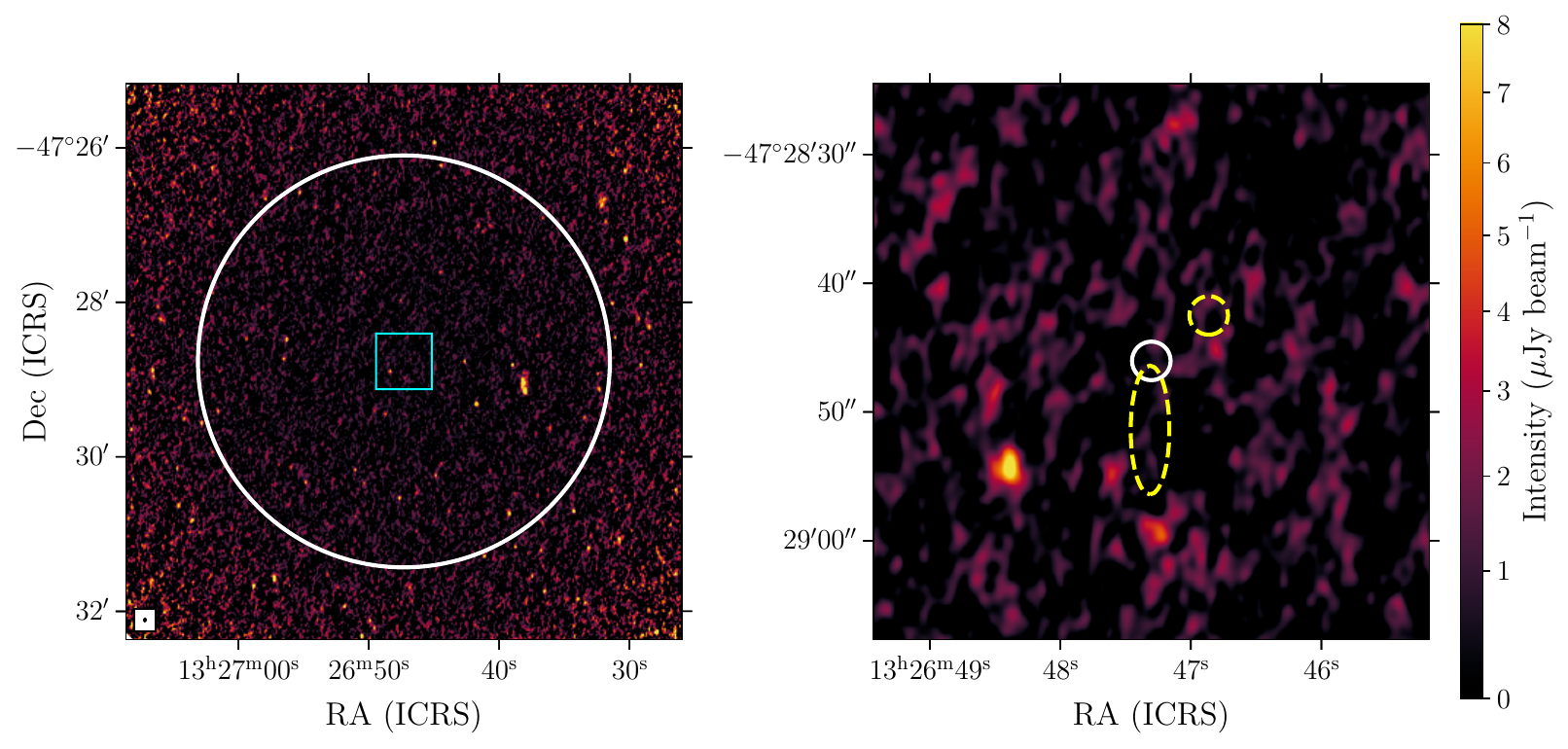}
\caption{The image on the left shows the core region of \om~(white circle). The zoomed-in region in the right image is indicated by the cyan box. The right image shows the recent centers proposed by \cite{2024MNRAS.528.4941P} (yellow) and \cite{2024Natur.631..285H} (white) for \om\ (Table~\ref{tab:centers}).}
\label{fig:omegacen}

\end{figure*}

We did not detect radio sources at any of these locations, with a 3$\sigma$ upper limit of $S_{\nu,\mathrm{lim}} = 3.3~\mu$Jy beam$^{-1}$ within the reported uncertainties of any of the proposed photometric centers listed in Table~\ref{tab:centers}. Furthermore, our reported $3\sigma$ upper limit is measured from a circular region with a $\sim$20$^{\prime\prime}$ radius within the cluster core, which is free from any bright sources. The closest 5-sigma unidentified source is $\sim$13$^{\prime\prime}$ away from the Center B. The deepest view of the core of \om\ based on our 7.25 GHz image is shown in Figure~\ref{fig:omegacen} (left), with the proposed cluster center locations (right). This non-detection places stringent constraints on the presence of a central radio-emitting source, limiting potential accretion signatures from an IMBH in the cluster core. 

\begin{table}[ht!]
    \centering
    \caption{The most recent photometric centers and their uncertainties in arcsec from the literature.}
    \label{tab:centers}
    \begin{tabular}{ccc}
    \hline
    \hline
    \textbf{Center} & \textbf{RA(J2000)} &\textbf{DEC(J2000)}\\ 
       & (hr:min:sec)  ($^{\prime\prime}$) &(\degree:$^{\prime}$:$^{\prime\prime}$) ($^{\prime\prime}$)  \\
    \hline
    \hline

    A       & 13:26:47.31 (1.5)  & -47:28:51.40 (5) \\
    B       & 13:26:46.86 (1.5)  & -47:28:42.50 (1.5)\\
    C       & 13:26:47.30 (0.6)  & -47:28:46.03 (0.4)\\
    \hline
    \hline
    \end{tabular}
    \tablecomments{\om\ center coordinates labeled as A and B are estimated locations based on kinemetry and velocity dispersion, respectively \citep{2024MNRAS.528.4941P}. Center C is the proposed IMBH location by \cite{2024Natur.631..285H} based on acceleration measurements. We detect no significant radio sources at any of these locations.}
\end{table}

\section{Analysis and discussion} \label{sec:diss}

We report on radio non-detections from the center of \om, in an ultra-deep (172\,hours) radio continuum image. The flux density upper limit from the deepest stacked image corresponds to a radio luminosity of $L_{\mathrm{R}} \leq 6.55 \times 10^{26}$~erg\,s$^{-1}$ at the kinematic distance of \om~($d = 5494\pm61$\,pc; \citealt{2025ApJ...983...95H}). The luminosity was derived using 
\begin{equation}
    L_{\mathrm{R}} = 4\pi d^{2} S_{\nu} \, \nu,
\end{equation}
where $S_{\nu}$ is the upper limit on the flux density, $d$ is the source distance, and $\nu$ is the observing frequency. We adopted a $3\sigma_{\mathrm{rms}}$ upper limit from the stacked image, corresponding to $S_{\nu,\mathrm{lim}} = 3.3~\mu$Jy at $\nu = 5.5$~GHz. 
Our findings suggest that either the inferred IMBH at the center of \om~must be residing in a low-density environment or that the accretion process from the intra-cluster medium is extremely inefficient. The interferometer spatially filters out diffuse Galactic emission. The field is not crowded, and no compact sources are detected within the synthesised beam around the target central position in either the individual or stacked images in both frequency bands. In the following sections, we discuss these implications.

\subsection{Constraints accretion efficiency from the fundamental plane}

The fundamental plane equation with black hole mass as a dependent variable (e.g., \citealt{2012ApJ...755L...1M, 2024ApJ...961...54P}) is expressed as \( \log M_{BH} = (1.638 \pm 0.070) \log L_R - (1.136 \pm 0.077) \log L_X - (6.863 \pm 0.790)\), where $M_{\rm BH}$ is the mass of the black hole, $L_R$ is radio luminosity and $L_X$ is X-ray luminosity. Although, following \cite{2012ApJ...750L..27S,2018ApJ...862...16T,  2024ApJ...961...54P}, we can assume that the accretion rate is a fraction $F=3$\% of the Bondi rate \citep{1952MNRAS.112..195B,2005ApJ...624..155P} and that the efficiency scales with mass accretion rate below 2 \% of Eddington, to eliminate $L_X$ in favor of \(M_{BH}\) and the intracluster gas density. The radio luminosity can be expressed as $\log L_R = \log(4\pi\nu) +2\log d + \log S_{\nu}$,  where $d$ is the distance, $\nu$ is the observed frequency, and $S_{\nu}$ is the flux density. We assume that the X-ray luminosity can be derived from the standard equation \(L_X = \epsilon \dot{M} c^{2}\), where $\dot{M}$ is the mass accretion rate, $c$ is the speed of light, and $\epsilon$ is the accretion efficiency. The X-ray component can then be expressed as $\log L_X = 3\log M_{BH} + \log A$ which gives the following form of the fundamental plane equation \citep{PhDPaduano2023}:


\begin{align}
\log M_{BH} &= 0.37\log(4\pi\nu) + 0.74\log d \nonumber \\ 
            &\quad + 0.37\log S_{\nu} - 0.257\log A - 1.53,
\label{eq:fpequation}
\end{align}

\noindent where $A$ (erg\,s$^{-1}$) is a scaling factor that sets the proportionality between black hole mass and X-ray luminosity. More detailed information can be found in the cited references.

This scaling factor can be expressed as
\begin{align}
    A = 0.125 (\epsilon \rho^2 F^2) \frac{ \pi K^2 c^3 G^3 M_\odot^3 \sigma_T }{ m_p c_s^6 },
\label{eq:A}
\end{align}

where \( \epsilon \) is the accretion efficiency \( \rho \) is the gas density and $F$ is the fraction of the Bondi rate at which accretion proceeds, which as stated above we took to be 0.03. The remaining quantities are physical constants: \( K \) is a dimensionless adiabatic constant determined by the thermodynamic regime (i.e., \( K = e^{3/2} \) for isothermal gas, \( K = 1 \) for adiabatic gas), \( G \) is the gravitational constant, \( M_\odot \) is the solar mass, \( \sigma_T \) is the Thomson cross-section, and \( m_p \) is the proton mass. The sound speed \( c_s \) is given by \(c_s = \sqrt{ \frac{ \gamma k_B T }{ \mu m_p } }\), where \( \gamma \) is the adiabatic index (\( \gamma = 1 \) for isothermal and \( \gamma = 5/3 \) for adiabatic gas), \( k_B \) is Boltzmann’s constant, \( T \) is the gas temperature, and \( \mu \) is the mean molecular weight.

High-precision HST proper-motion measurements of fast-moving stars within the inner \(10^{\prime\prime}\) region of \om were compared with state-of-the-art N-body simulations for various IMBH masses. The mass estimations depend on the adopted dynamical model. The velocity measurements alone provide a firm lower limit of $8200\,M_{\odot}$, while models incorporating the full velocity distribution favours a central IMBH with a mass in the range $39000 - 47000\,M_{\odot}$ \citep{2024Natur.631..285H}. Using this most robust mass estimate together with our deepest radio upper limit, we can refine our understanding of the accretion efficiency of the IMBH in \om.

Using Equations~\ref{eq:fpequation} and \ref{eq:A}, we can estimate the accretion efficiency, assuming that the ambient gas density in \om\ is similar to that measured in other globular clusters. The gas density is defined as \(\rho = n\mu m_H\), where \(n\) is the gas number density and \(\mu\) is the mean molecular weight, consistent with the sound speed. Following \citet{2012ApJ...750L..27S}, we assume \(n = 0.2\pm0.1 \, \text{cm}^{-3}\), based on the estimated free electron density derived using pulsars in other globular clusters \citep{1992PhDT........20A,2001ApJ...557L.105F, 2025SciBu..70.1568Z}. 
With these considerations, given the estimated mass of the black hole, our radio upper-limit yields 99\% upper limits on the accretion efficiency of \(\epsilon \lesssim 8\times10^{-5} \) (adiabatic) and \(\epsilon \lesssim 9\times10^{-7} \) (isothermal) assuming the Bondi fraction \(F=0.03\). Conservatively, accounting for the scatter in the fundamental plane \citep[e.g.,][]{2019ApJ...871...80G} increases these upper limits to $ 4\times10^{-3}$ and $5\times10^{-5}$ for the adiabatic and isothermal cases, respectively. These results provide the deepest constraints to date on the accretion efficiency of an IMBH.  For comparison, previous studies have reported accretion efficiency constraints for a stellar-mass black hole in NGC~3201 as $\epsilon \leq 1.5 \times 10^{-5}$ (for system ACS ID \#21859; \citealt{2022MNRAS.510.3658P}), for Gaia~BH1 as $\epsilon \leq 10^{-6} - 10^{-5}$, and for Gaia~BH2 as $\epsilon \leq 10^{-4} - 10^{-3}$ (\citealt{2024PASP..136b4203R}). 

The ratio of radio luminosity to Eddington luminosity, \( L_R / L_{\mathrm{Edd}} \), provides a scale-independent approach to assess black hole activity. Using the upper limits of radio luminosity (\( L_R \leq 6.55 \times 10^{26} \, \mathrm{erg\,s^{-1}} \)) derived from our data along with the black hole mass limits $39000 - 47000\,M_{\odot}$ gives \( L_R / L_{\mathrm{Edd}} \lesssim (1.3 - 1.1)\times 10^{-16} \). \citet{2023Galax..11...53Y} studied radio and X-ray emissions from a sample of black holes, which also included the largest sample of IMBH candidates. They have reported that only a small fraction ($\sim 0.6\%$ ) of their IMBH candidate-hosted galaxies are active in radio, suggesting a long quiescent phase in the evolution of IMBHs. Comparatively, two of the low accreting IMBHs from \citet{2023Galax..11...53Y}, namely, NGC 4394 \citep{2012ApJ...753..103N,2020MNRAS.496.4061D} and NGC 404 \citep{2001ApJS..133...77H,2006ApJ...646L..95W}, have \(L_R / L_{\mathrm{Edd}} \lesssim 10^{-10} -  10^{-9} \) \citep{2023Galax..11...53Y}. This comparison indicates a similar possibility of environmental effects on accretion in \om\,cluster core. Recent simulations show that IMBH accretion rates depend strongly on the cluster environment, evolutionary timescale, gas supply, and growth of the putative IMBH \citep{2023MNRAS.521.2930R,2025MNRAS.537..956P,2025A&A...700A..52W}.

The expected mean-squared displacement of an IMBH within the cluster can be expressed as $ \langle x^2 \rangle = \frac{2}{9} \left(\frac{m_\star}{M_\text{BH}}\right) r_c^2$, where \( m_\star \) is the average stellar mass in the cluster core, \( M_\text{BH} \) is the IMBH mass, and \( r_c \) is the core radius \citep{2002ApJ...572..371C}. Adopting a conservative estimate of \( m_\star = 1 M_{\odot} \), due to core depletion from mass segregation in \om, we find an expected displacement range of \( 0.38^{\prime\prime} - 0.34^{\prime\prime}\) for the mass range of $39000 - 47000\,M_{\odot}$ from \citet{2024Natur.631..285H}, which is approximately \( 0.2\% \) of the core radius (\( 160^{\prime\prime} \); \citealt{2018MNRAS.478.1520B}).


\section{Summary and Conclusion} \label{sec:concl}

The debate surrounding the potential existence of an IMBH in \om~has persisted for over a decade, supported by theoretical predictions and observational evidence. The recent study of \citet{2024Natur.631..285H} resolved this debate, using fast-moving stars at the center of \om~ to verify the presence of an IMBH, with an estimated mass of 39,000--47,000$M_{\odot}$. By including radio observations, which are sensitive to emission that may arise from accretion processes associated with a central black hole, we can probe the accretion efficiency of this IMBH.

We present the most sensitive radio observations of \om~to date, achieving an RMS noise of $1.1~\mu$Jy~beam$^{-1}$ at $7.25$ GHz and a $3\sigma$ upper limit on the radio luminosity of $L_R \leq 6.55 \times 10^{26}$ erg\,s$^{-1}$. On combining this radio limit with the stringent mass estimate of $39000 - 47000\,M_{\odot}$ from \citet{2024Natur.631..285H}, 
it suggests that the IMBH in $\omega$ must be accreting at an exceptionally low rate, with a conservative $3\sigma$ upper limit of $\epsilon \leq 4 \times 10^{-6}$. Alternatively, the absence of detectable emission may imply that the gas density is too low to produce accretion.

Future facilities such as the Square Kilometre Array (SKA)\footnote{\url{https://www.skao.int/en}} or the next-generation Very Large Array (ngVLA)\footnote{\url{https://ngvla.nrao.edu/}} could provide even deeper constraints on the nature of an IMBH in \om; for example, to reach our sensitivity level using the SKA-mid AA4 array configuration with Briggs weighting would only require $\sim$15 minutes of observation time.

Beyond searching for the accretion signature of an IMBH, our deep radio image will allow us to produce a catalogue of all radio sources within the half-light radius of \om, which can be cross-matched with deep images from other wavelengths \citep[e.g.,][]{2018MNRAS.479.2834H}. We plan to analyse all detected radio sources in the cluster in a future work.

\begin{acknowledgments}
The authors would like to thank the anonymous reviewer for helpful comments on the manuscript. ADM would like to thank Thomas Crowther for his support in formatting Figure \ref{fig:omegacen}. The authors would like to thank J. Stevens for assisting in scheduling the observations. The Australia Telescope Compact Array is part of the Australia Telescope National Facility (grid.421683.a) which is funded by the Australian Government for operation as a National Facility managed by CSIRO. We acknowledge the Gomeroi people as the traditional owners of the Observatory site. The authors acknowledge extensive use of NASA’s Astrophysics Data System Bibliographic Services and arXiv.
\end{acknowledgments}

\vspace{5mm}
\facilities{Australia Telescope Compact Array}


\software {Miriad \citep{1995ASPC...77..433S},  
           CASA \citep{2022PASP..134k4501C}, 
           CARTA \citep{2021zndo...4905459C}
           Astropy \citep{2018AJ....156..123A}, 
           Matplotlib \citep{2007CSE.....9...90H},
           Numpy \citep{2011CSE....13b..22V}
          }


\bibliography{papers}{}
\bibliographystyle{aasjournal}



\end{document}